   \newcommand{\be}[0]{\begin{equation}}
   \newcommand{\ee}[0]{\end{equation}}
   \newcommand{\ba}[0]{\begin{eqnarray}}
   \newcommand{\ea}[0]{\end{eqnarray}}
\documentclass[12pt]{article}
\usepackage{epsfig} \usepackage{amssymb} \usepackage{amsfonts}
\begin{document}
\Large
\hfill\vbox{\hbox{IPPP/02/21}
            \hbox{DTP/02/42}
            \hbox{REVISED June 2002}}
\nopagebreak

\vspace{0.75cm}
\begin{center}
\LARGE
{\bf All-orders infra-red freezing of ${R}_{{e}^{+}{e}^{-}}$ in perturbative QCD}
\vspace{0.6cm}
\Large

D.~M.~Howe\footnote{email:{\tt d.m.howe@durham.ac.uk}} and C.~J.~Maxwell\footnote{email:{\tt c.j.maxwell@durham.ac.uk}}

\vspace{0.4cm}
\large
\begin{em}
Centre for Particle Theory, University of Durham\\
South Road, Durham, DH1 3LE, England
\end{em}

\vspace{1.7cm}

\end{center}
\normalsize
\vspace{0.45cm}

\centerline{\bf Abstract}
\vspace{0.3cm}
%%%%%%%%%%%%ABSTRACT%%%%%%%%%%%%%%%%%%
We consider the behaviour of the perturbative QCD corrections to
the $R_{{e^+}{e^-}}$ ratio, in the limit that the c.m. energy
$\sqrt{s}$ vanishes. Writing $R_{{e^+}{e^-}}(s)=3{\sum_{f}}{Q}_{f}^{2}(1+{\cal{R}}(s))$,
with $Q_f$ denoting the electric charge of quark flavour $f$, we find that
for ${N_f}<9$ flavours of massless quarks, the perturbative correction ${\cal{R}}(s)$ to the
parton model result
smoothly approaches from below the infra-red limit ${\cal{R}}(0)=2/b$, as $s{\rightarrow}0$.
Here $b=(33-2{N_f})/6$ is the first QCD beta-function coefficient. This freezing holds
to all-orders in perturbation theory. The $s$-dependence can be written analytically
in closed form in terms of the Lambert $W$
function.

\newpage
In the ultra-violet ${s}{\rightarrow}{\infty}$ limit of QCD the renormalized coupling
vanishes, and this property of Asymptotic Freedom
 underwrites the successful use of perturbative methods
in testing the theory \cite{r1}. In the infra-red limit $s{\rightarrow}0$, however,
one may expect that perturbation theory will break down , with typically a Landau pole
singularity in the coupling when $s{\sim}{\Lambda}_{QCD}^{2}$, and that non-perturbative effects
will be important. However, the phenomenological virtues of assuming a frozen couplant,
with the renormalized ${\alpha}_{s}(s)$ approaching a constant value ${\alpha}_{s}(0)/{\pi}\sim{0.3}$
in the infra-red have long been recognised [2-5]. In a pioneering paper Mattingly and
Stevenson investigated the behaviour of the perturbative corrections to ${R}_{{e^+}{e^-}}$
including third-order QCD corrections,
in the framework of the Principle of Minimal
Sensitivity (PMS) approach \cite{r2}. Their PMS optimized coupling indeed froze to a value
around $0.26$ below $300$ MeV. These predictions were then smeared using the technique
of Poggio-Quinn-Weinberg (PQW) \cite{r5}, and were in suprisingly good agreement with
similarly smeared experimental data for ${R}_{{e^+}{e^-}}$.
Some scepticism about the existence of infra-red fixed point behaviour had previously been
expressed \cite{r6}.
 In this letter we wish to demonstrate that including all-orders
in perturbation theory the perturbative corrections to ${R}_{{e^+}{e^-}}$ do freeze
in the infra-red. The limiting value being $2/b$, where $b=(33-2{N_f})/6$ is the first
beta-function coefficient of QCD with $N_f$ quark flavours. We assumed massless quarks
and for freezing to this limit one requires ${N_f}<9$ flavours. In fact the freezing
behaviour does not correspond to an infra-red fixed point in the beta-function, but
rather stems from the energy dependence induced by analytical continuation from the Euclidean to
Minkowskian region in defining ${R}_{{e}^{+}{e}^{-}}$.\\

We begin by defining the ${R}_{{e^+}{e^-}}$ ratio at c.m. energy $\sqrt{s}$,
\be
{R}_{{e^+}{e^-}}(s){\equiv}\frac{{\sigma}
_{tot}({e}^{+}{e}^{-}\rightarrow\rm{hadrons})}{{\sigma}(
{e}^{+}{e}^{-}\rightarrow{\mu}^{+}{\mu}^{-})}=3{\sum_{f}}{Q}_{f}^{2}(1+{\cal{R}}(s))\;.
\ee
Here the $Q_f$ denote the electric charges of the different flavours of quarks, and ${\cal{R}}(s)$
denotes the perturbative corrections to the parton model result, and has a perturbation series
of the form,
\be
{\cal{R}}(s)=a+
{\sum_{n>0}}{r}_{n}{a}^{n+1}\;.
\ee
Here $a{\equiv}{\alpha}_{s}({\mu}^{2}
)/{\pi}$ is the renormalized coupling, and the
coefficients $r_1$ and $r_2$ have been computed in the $\overline{MS}$ scheme with
renormalization scale ${\mu}^{2}=s$ \cite{r7,r8}.
${R}_{{e}^{+}{e}^{-}}$ is directly related
to the transverse part of the correlator of two vector currents in the Euclidean region,
\be
({q}_{\mu}{q}_{\nu}-{g}_{{\mu}{\nu}}{q}^{2}){\Pi}(s)=4{\pi}^{2}i{\int}{d}^{4}x{e}^{iq.x}<0|T[{j}_
{\mu}(x){j}_{\nu}(0)]|0>\;,
\ee
where $s=-{q}^{2}>0$. To avoid an unspecified constant it is convenient to take a
logarithmic derivative with respect to $s$ and define the Adler $D$-function,
\be
{D}(s)=
-s\frac{d}{ds}{\Pi}(s)\;.
\ee
This can be represented by Eq.(1) with the perturbative corrections ${\cal{R}}(s)$ replaced by
\be
{\cal{D}}(s)=a+
{\sum_{n>0}}{d}_{n}{a}^{n+1}\;.
\ee
 The Minkowskian observable ${\cal{R}}(s)$ is related to
${\cal{D}}(-s)$ by analytical continuation from Euclidean to Minkowskian. This can be
elegantly formulated as an integration around a circular contour in the complex
energy-squared $s$-plane \cite{r9,r10},
\be
{\cal{R}}(s)=\frac{1}{2{\pi}}{\int_{-{\pi}}^{\pi}}{d{\theta}}{\cal{D}}(s{e}^{i{\theta}})\;.
\ee
Expanding ${\cal{D}}(s{e}^{i{\theta}})$ as a power series in $\bar{a}{\equiv}{a}(s{e}^{i{\theta}})$,
and performing the ${\theta}$ integration term-by-term, leads to a ``contour-improved''
perturbation series, in which at each order an infinite subset of analytical continuation
terms present in the conventional perturbation series of Eq.(2) are resummed. It is this
complete analytical continuation that builds our claimed freezing of ${\cal{R}}(s)$. In
contrast Ref.\cite{r2} used the conventional fixed-order perturbative expansion of Eq.(2) in
the PMS approach. It is useful to begin by considering the ``contour-improved'' series for
the simplified case of a one-loop coupling. The one-loop coupling will be given by
\be
a(s)=\frac{2}
{b{\ln}(s/{\tilde{\Lambda}}_{\overline{MS}}^{2})}\;.
\ee
As described above one can then obtain the ``contour-improved'' perturbation series
for ${\cal{R}}(s)$,
\be
{\cal{R}}(s)={A}_{1}(s)+{\sum_{n=1}^{\infty}}{d}_{n}{A}_{n+1}(s)\;,
\ee
where the functions ${A}_{n}(s)$ are defined by,
\ba
{A}_{n}(s)&{\equiv}&\frac{1}{2{\pi}}{\int_{-{\pi}}^{\pi}}{d{\theta}}{\bar{a}}^{n}=
\frac{1}{2{\pi}}{\int_{-\pi}^{\pi}}
{d{\theta}}\frac{{a}^{n}(s)}{{[1+ib{\theta}a(s)/2
]}^{n}}\;.
\ea
This is an elementary integral which can be evaluated in closed-form as \cite{r11a}
\ba
{A}_{1}(s)&=&\frac{2}{{\pi}b}{\arctan}\left(\frac{{\pi}ba(s)}{2}\right)
\nonumber \\
{A}_{n}(s)&=&
\frac{2{a}^{n-1}(s)}{b{\pi}(1-n)}
Im\left[{\left(1+
\frac{ib{\pi}a(s)}{2}\right)}^{1-n}
\right]\;(n>1)\;.
\ea
We then obtain the one-loop ``contour-improved'' series for ${\cal{R}}(s)$,
\be
{\cal{R}}(s)=\frac{2}{\pi{b}}{\arctan}\left(\frac{\pi{b}{a}(s)}{2}\right)+
{d}_{1}\left[\frac{{a}^{2}(s)}{(1+{b}^{2}{\pi}^{2}{a}^{2}(s)/4)}\right]
+{d}_{2}\left[\frac{{a}^{3}(s)}{{(1+{b}^{2}{\pi}^{2}{a}^{2}(s)/4)}^{2}}\right]+\ldots\;.
\ee
The first $\arctan$ term is well-known, and corresponds to resumming the infinite
subset of analytical continuation terms in the standard perturbation series of
Eq.(2) which are independent of the $d_n$ coefficients. Subsequent terms
corrrespond to resumming to all-orders the infinite subset of terms in Eq.(2)
proportional to ${d}_{1},{d}_{2},{\ldots}$, etc. It is crucial to notice that
in each case the resummation is {\it convergent}, provided that $|a(s)|<2/{\pi}{b}$.
Thus, even though the series in Eqs.(2,5) are divergent because of the $n!$ growth
of the $d_n$ coefficients,
the resummations implicit in performing the integration
of Eq.(6) term-by-term are well-defined. In the ultra-violet $s\rightarrow{\infty}$ limit
the ${A}_{n}(s)$ vanish as required by asymptotic freedom. In the infra-red
$s\rightarrow{0}$ limit, the one-loop coupling $a(s)$ has a ``Landau'' singularity
at $s={\tilde{\Lambda}}_{\overline{MS}}^{2}$.
 However, the functions ${A}_{n}(s)$
resulting from resummation, if analytically continued,
 are well-defined for all real values of $s$. ${A}_{1}(s)$
smoothly approaches from below the asymptotic infra-red value $2/b$, whilst for
$n>1$ the ${A}_{n}(s)$ vanish.
 Thus, as claimed,
 ${\cal{R}}(s)$ is asymptotic to
$2/b$ to all-orders in perturbation theory. It might be thought that the existence
of such an infra-red limit,
independent of the higher-order structure of perturbation theory, is remarkable.
 In fact, however, it is to be expected. A useful analogy is
the exponentiation to all-orders of large infra-red logarithms which appear for
jet observables such as thrust distributions in ${e}^{+}{e}^{-}$ annihilation \cite{r1}. Here
the standard fixed-order
perturbation theory breaks down in the two-jet region as
these logarithms become infinite. However, if the large logarithms are resummed
to all-orders one builds an exponential factor , and the thrust distribution
smoothly approaches zero in the two-jet region,
to all-orders in perturbation
theory. We should note a subtlety in our derivation above. As defined by the
integration in $\theta$ around the circle in Eq.(9)
the ${A}_{n}(s)$ are not
defined for $s\leq
{\tilde{\Lambda}}_{\overline{MS}}^{2}$
. However the ${A}_{n}(s)$
given by Eq.(10)
are defined for all real $s$. A more careful derivation would use
instead of the contour integral as a starting point, the dispersion relation,
\be
{\cal{R}}(s)=\frac{1}{2\pi{i}}{\int_{-s-i{\epsilon}}^{-s+i{\epsilon}}}{dt}\frac{{\cal{D}}(t)}{t}\;,
\ee
which clearly avoids the ``Landau'' singularity and is well-defined for all real $s$.
One can directly obtain the result of Eq.(10) from the dispersion relation by
simple manipulations. It should be noted that this route is equivalent to the
Analytic Perturbation Theory (APT) approach, in which it is advocated that Minkowskian
observables are expanded in a basis of functions ${\mathfrak{A}}_{n}(s)$ obtained by
performing an integral transform of the Euclidean coupling using the dispersion
relation of Eq.(12). In the case of the ${R}_{{e}^{+}{e}^{-}}$ ratio the resulting
APT expansion corresponds to the ``contour-improved'' expansion of Eq.(8) and
${\mathfrak{A}}_{n}(s)={A}_{n}(s)$, given by Eq.(10). The infra-red freezing
and absence of a ``Landau pole'' in the ${\mathfrak{A}}_{n}(s)$ has previously
been discussed in the APT approach, and provides one of its major motivations.
Analytic expressions for the one-loop ${\mathfrak{A}}_{n}(s)$ have been given, and
freezing to all-loops in APT can be demonstrated.
For a recent review see Ref.\cite{r11b},
and references therein.\\

We now move to the more challenging problem of what happens for realistic
QCD beyond the simple one-loop approximation.
The freezing is most easily analysed using a renormalization scheme in which the
beta-function equation has its two-loop form,
\be
\frac{\partial{a}({\mu}^{2})}
{\partial{\ln}{\mu}^{2}
}=-\frac{b}{2}
{a}^{2}({\mu}^{2})(1+c{a}({\mu}^{2}))\;.
\ee
This corresponds to a so-called 't Hooft scheme \cite{r11} in which the non-universal
beta-function coefficients are all zero. Here $c=(153-19{N_f})/12b$ is the
second universal beta-function coefficient. The key feature of these schemes is
that the coupling can be expressed analytically in closed-form in terms of the
Lambert $W$ function
, defined implicitly by $W(z){\exp}(W(z))=z$ \cite{r12}. One has
\ba
{a}({\mu}^{2})
&=&-\frac{1}{c[1+{W}_{-1}(z(\mu))]}
\nonumber \\
z(\mu)&{\equiv}&-\frac{1}{e}{\left(\frac{\mu}{{\tilde{\Lambda}}_{\overline{MS}}}\right)}^{-b/c}\;,
\ea
where ${\tilde{\Lambda}}_{\overline{MS}}$ is defined according to the convention of \cite{r13}
, and is related to the standard definition \cite{r14} by ${\tilde{\Lambda}}_{\overline{MS}}=
{(2c/b)}^{-c/b}
{\Lambda}_{\overline{MS}}$. The ``$-1$'' subscript on $W$ denotes the
branch of the Lambert $W$ function required for Asymptotic Freedom, the nomenclature being that of Ref.\cite{r15}. Assuming
a choice
of renormalization scale ${\mu}^{2}=xs$ , where $x$ is a dimensionless constant, for
the perturbation series of ${\cal{D}}(s)$ in Eq.(5), one can then expand
the integrand in Eq.(6) for ${\cal{R}}(s)$ in powers of ${\bar{a}}\equiv{a}(xs{e}^{i{\theta}})$
, which can be expressed in terms of the Lambert $W$ function using Eq.(14),
\be
{\bar{a}}=\frac{-1}{c[1+W(A(s){e}^{iK{\theta}})]}
\ee
where
\be
A(s)=\frac{-1}{e}{\left(\frac{\sqrt{xs}}{{\tilde{\Lambda}}_{\overline{MS}}}\right)}^{-{b/c}}\;\;\;,
{K}=\frac{-b}{2c}\;.
\ee
The functions ${A}_{n}(s)$ in the ``contour-improved'' series
 are then given, using Eqs(15,16), by
\ba
{A}_{n}(s)&{\equiv}&\frac{1}{2{\pi}}{\int_{-{\pi}}^{\pi}}{d{\theta}}{\bar{a}}^{n}=
\frac{1}{2{\pi}}{\int_{-\pi}^{0}}{d{\theta}}\frac{{(-1)}^{n}}{c^n}{[1+{W}_{1}(A(s){e}^{iK{\theta}})]}^{-n}
\nonumber \\
&+&\frac{1}{2\pi}{\int_{0}^{\pi}}{d{\theta}}\frac{{(-1)}^{n}}{c^n}{[1+{W}_{-1}(A(s){e}^{iK{\theta}})]}^{-n}\;.
\ea
Here the appropriate branches of the $W$ function are used in the two regions of integration. As discussed
in Refs.\cite{r16,r16a},
by making the change of variable $w=W(A(s){e}^{iK{\theta}})$ we can then obtain
\be
{A}_{n}(s)=\frac{{(-1)}^{n}}{2iK{c^n}{\pi}}{\int^{{W}_{-1}(A(s){e}^{iK\pi})}_{{W}_{1}(A(s){e}^{-iK\pi})}}
\frac{dw}{w{(1+w)}^{n-1}}\;.
\ee
This is an elementary integral and noting that ${W}_{1}(A(s){e}^{-iK\pi})={[{W}_{-1}(A(s){e}^{iK\pi})]}^{\ast}$, we
obtain for $n=1$,
\be
A_{1}(s)=\frac{2}{b}-\frac{1}{\pi{K}c}Im[{\ln}({W}_{-1}(A(s){e}^{iK\pi}))]\;,
\ee
where the $2/b$ term is the residue of the pole at $w=0$. In Ref.\cite{r16} this
contribution was omitted in error. For $n>1$ we obtain
\be
{A}_{n}(s)=\frac{{(-1)}^{n}}{{c^n}K{\pi}}Im\left[{\ln}\left(\frac{{W}_{-1}(A(s){e}^{iK\pi})}{1+{W}_{-1}(A(s){e}^{iK\pi})}\right)
+{\sum_{k=1}^{n-2}}\frac{1}{k{(1+{W}_{-1}(A(s){e}^{iK\pi}))}^k}\right]\;.
\ee
Here the contributions from the poles at $w=0$ and $w=-1$ cancel exactly.
Equivalent expressions for the ${\mathfrak{A}}_{n}(s)$ have been
obtained in the APT approach \cite{r16a}.
 Provided that
${b/c}>0$, which will be true for ${N_f}<9$, the
functions ${A}_{n}(s)$ are well-defined for all real values of $s$.
As $s\rightarrow{\infty}$ they vanish
as required by Asymptotic Freedom. As $s\rightarrow{0}$,
$A_{1}(s)$ smoothly approaches the infra-red
limit $2/b$ from below, as the second term in Eq.(19) vanishes in the limit. Whilst
for $n>1$
the ${A}_{n}(s)$ vanish as $s\rightarrow{0}$, and so, as in the one-loop case,
${\cal{R}}(s)$ is asymptotic to $2/b$ to all-orders in perturbation theory. The
cancellation of pole contributions noted above is crucial in achieving this.
We should point out that for ${b/c}>4$ we move
to other branches of the Lambert $W$
function in order
to keep ${a}({\mu}^{2})$ continuous. This just changes the value of the
branch of the Lambert $W$ function in ${A}_{n}(s)$, and will not alter our result for $s\rightarrow{0}$. We find ${b/c}>4$
first occurring for ${N}_{f} = 7$. Again
we need to
refine the above argument slightly, using the dispersion relation of Eq.(12), which after
a change of variable
 $w=W(A(t))$ yields the integral of Eq.(18). As in the one-loop case the ${A}_{n}(s)$ of
 Eqs.(19,20) can be obtained as all-orders resummations of subsets of terms in the standard
 perturbation series of Eq.(2), these resummations are convergent provided $|a(s)|<2/{\pi}b$.
 In Figures 1-3 we plot the functions ${A}_{1}(s),{A}_{2}(s)$ and ${A}_{3}(s)$, respectively,
as functions of $sx/{\tilde{\Lambda}}_{\overline{MS}}^{2}$. ${N_f}=2$ flavours of quark are assumed.
\nopagebreak
\begin{figure}
\begin{center}
\epsfig{file=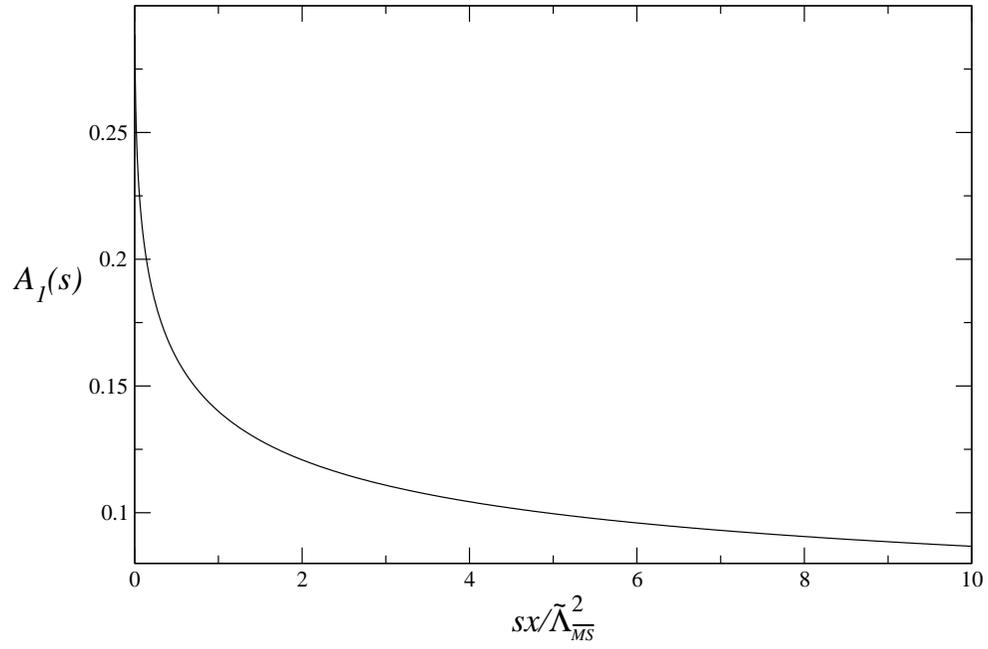,angle=270,width=15cm}
\caption{The function ${A}_{1}(s)$ of Eq.(19)
versus $sx/{\tilde{\Lambda}}_{\overline{MS}}^{2}$. We assume
${N_f}=2$ flavours of quark.}

\label{fig:f1}
\end{center}
\end{figure}
\begin{figure}
\begin{center}
\epsfig{file=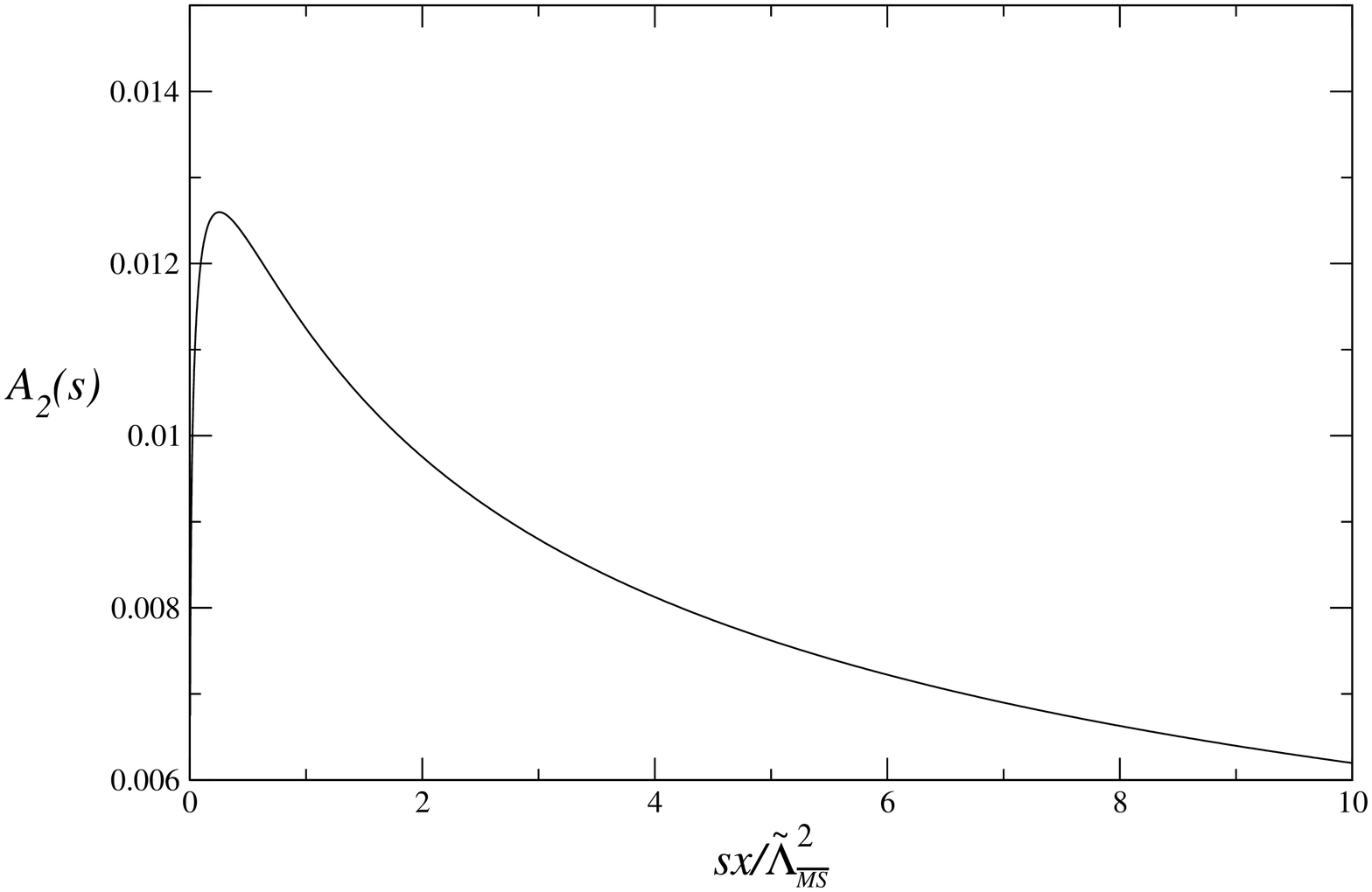,angle=270,width=15cm}
\caption{As Fig.1 but for ${A}_{2}(s)$ of Eq.(20).}
\label{fig:f2}
\end{center}
\end{figure}
\newpage
\begin{figure}
\begin{center}
\epsfig{file=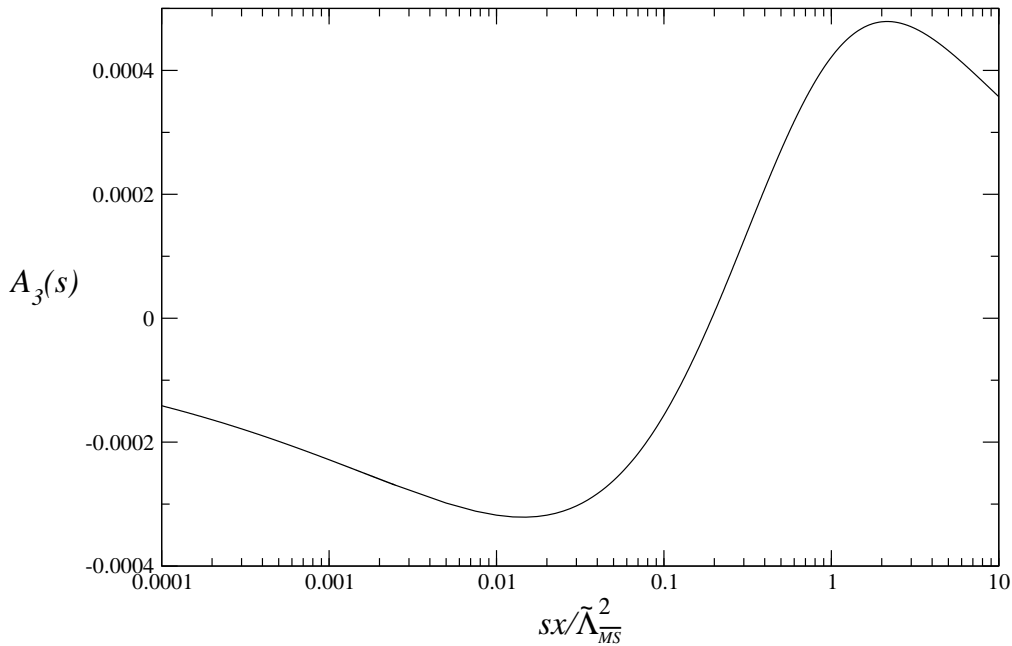,angle=270,width=15cm}
\caption{As Fig.1 but for ${A}_{3}(s)$ of Eq.(20).}
\label{fig:f3}
\end{center}
\end{figure}
 \nopagebreak
We have shown that the freezing occurs to all-orders in perturbation theory, and thus must
occur independent of the choice of renormalization scheme. The use of a 't Hooft scheme turns
out to make the freezing manifest. We used a general choice of renormalization scale
${\mu}^{2}=xs$. The true infra-red $s$-dependence of ${\cal{R}}(s)$ does not depend on
the unphysical parameter $x$. One should rather eliminate the ${\mu}$-dependence of the
result altogether by completely resumming all the ultra-violet logarithms which build
the $s$-dependence. This so-called Complete Renormalization Group Improvement (CORGI)
approach \cite{r17} corresponds to choosing ${\mu}^{2}={e}^{-2d/b}s$, where $d$ is the
NLO perturbative correction $d_1$ to ${\cal{D}}(s)$ in Eq.(5), in the $\overline{MS}$ scheme
with ${\mu}^{2}=s$. One then has the ``contour-improved'' CORGI series,
\be
{\cal{R}}(s)={A}_{1}(s)+{\sum_{n=2}^{\infty}}{X}_{n}{A}_{n+1}(s)\;,
\ee
where the $X_n$ are the CORGI invariants, and only $X_2$ is known. Now
$A(s)=(-1/e){({\sqrt{s}}/{\Lambda}_{D})}^{-b/c}$, where ${\Lambda}_{D}{\equiv}{e}^{d/b}{\tilde{\Lambda}}_{\overline{MS}}$.
For an infra-red fixed point at ${\cal{R}}(0)={\cal{R}}^{\ast}$, corresponding to a zero of the beta-function,
 one expects the asymptotic
behaviour \cite{r2}
\be
{\cal{R}}(s)-{\cal{R}}^{\ast}{\sim}{s}^{\gamma}\;,
\ee
 where $\gamma$ is a critical exponent. Our freezing to $2/b$ is instead driven by the analytical continuation from
the Euclidean to Minkowskian regions, and one obtains the asymptotic behaviour,
\be
{\cal{R}}(s)-\frac{2}{b}{\approx}
\frac{-1/c-2/b}{{W}_{0}(-A(s))}\;.
\ee
\\
Again this involves the ubiquitous Lambert $W$ function.\\

We should stress that, of course, the result ${\cal{R}}(0)=2/b$ is in itself only of abstract
interest since $s=4{m}_{\pi}^{2}$ is the threshold in full QCD with massive quarks. The
important conclusion is that while the conventional perturbation series of Eq.(2)
breaks down at the spurious Landau pole
in the coupling $a(s)$, this is eliminated
by completely resumming all the analytical continuation terms, so that the ``contour-improved'' (or APT)
perturbation series is well-behaved in the infra-red. It was crucial in investigating
this to be able to evaluate the functions ${A}_{n}(s)$ in closed analytic form. In
previous phenomenological investigations \cite{r9,r10} these functions were evaluated
by numerical integration with Simpson's Rule, making it impossible to go further into
the infra-red than the Landau pole obstruction. Whilst the infra-red freezing is
most easily investigated using the ``contour improved'' or Analytic version of the
perturbation series in Eqs.(8,21), it should be stressed that the freezing is
an all-orders result of QCD perturbation theory. The standard expectation is that
by itself the all-orders perturbation series is ambiguous due to the presence
of infra-red renormalons, these renormalon ambiguities cancelling against corresponding
non-logarithmic UV divergences in the non-perturbative Operator Product Expansion (OPE)
\cite{r18}. The vanishing of the ${A}_{n}(s)$
for $n>1$ in the infra-red means
that these renormalon ambiguities also vanish, and so the
implication is that the non-logarithmic UV divergences of the OPE also vanish in the
infra-red. This of course says nothing about the infra-red limit of the resummed
OPE, indeed we know that there are important non-perturbative effects which build
a complicated set of hadronic resonances,
 but the interesting observation is that perturbative and non-perturbative
effects are {\it separately} well-defined in the infra-red limit for Minkowskian
quantities. This is not the case for Euclidean quantities where the Euclidean
APT coupling necessarily includes a resummation of non-perturbative OPE terms
\cite{r11b}.

There are evidently many phenomenological applications of the ``contour-improved''
CORGI perturbation series for ${\cal{R}}(s)$, in particular one can repeat the analysis
of Ref.\cite{r2} and compare PQW smeared \cite{r5} data for ${R}_{{e}^{+}{e}^{-}}(s)$
with the similarly smeared perturbative freezing. This exercise has been performed in
the APT approach in Ref.\cite{17a}, and good agreement found.
 There will also be applications
to estimating uncertainties in ${\alpha}(M_Z)$, and in estimating the hadronic
corrections to the anomalous magnetic moment of the muon. We hope to report on these
aspects in a future publication.

\section*{Acknowledgements}
We thank Dmitry Shirkov and Igor Solovtsov for helpful comments on an earlier
version of this paper.
D.M.H. gratefully acknowledges receipt of a PPARC UK Studentship.

\end{document}